\documentclass[10pt, conference, compsocconf]{IEEEtran}

\ifCLASSINFOpdf
  \usepackage[pdftex]{graphicx}
\else
  \usepackage[dvips]{graphicx}
  \usepackage{enumitem}
\fi

\usepackage[cmex10]{amsmath}

\usepackage{booktabs}
\usepackage{todonotes}
\usepackage{algpseudocode}
\usepackage{times}
\usepackage{fancyhdr,graphicx,amsmath,amssymb}
\usepackage[ruled,vlined]{algorithm2e}

\usepackage{array}

\usepackage{hyperref}

\usepackage{caption}
\let\oldbibliography\thebibliography
\renewcommand{\thebibliography}[1]{%
  \oldbibliography{#1}%
  \setlength{\itemsep}{0pt}%
  \setlength{\parskip}{0pt}%
}

\begin{document}
\bstctlcite{IEEEexample:BSTcontrol}
\title{ARKV: Adaptive and Resource-Efficient KV Cache Management under Limited Memory Budget for Long-Context Inference in LLMs}

\author{
    \IEEEauthorblockN{Jianlong Lei and Shashikant Ilager}
    \IEEEauthorblockA{
        Multiscale Networked Systems (MNS) Group, Informatics Institute \\
        University of Amsterdam, The Netherlands \\
        Email: leijianlong.jalon@gmail.com, s.s.ilager@uva.nl
    }
}

\maketitle

\begin{abstract}

Large Language Models (LLMs) are increasingly deployed in scenarios demanding ultra-long context reasoning, such as agentic workflows and deep research understanding. However, long-context inference is constrained by the KV cache, a transient memory structure that grows linearly with sequence length and batch size, quickly dominating GPU memory usage. Existing memory reduction techniques, including eviction and quantization, often rely on static heuristics and suffer from degraded quality under tight budgets. In this paper, we propose ARKV, a lightweight and adaptive framework that dynamically allocates precision levels to cached tokens based on per-layer attention dynamics and token-level importance. During a short prefill phase, ARKV estimates the original quantization (OQ) ratio of each layer by computing statistical scores such as attention entropy, variance and kurtosis. During decoding, tokens are assigned to one of three states—Original (full precision), Quantization (low precision), or Eviction—according to a fast heavy-hitter scoring strategy. Our experiments on LLaMA3 and Qwen3 models across diverse long- and short-context tasks demonstrate that ARKV preserves $\sim$97\% of baseline accuracy on long-context benchmarks while reducing KV memory usage by 4×, with minimal throughput loss. On short-context tasks, ARKV matches full-precision baselines; on GSM8K math reasoning, it significantly outperforms uniform quantization. These results highlight the practical viability of ARKV for scalable LLM deployment, offering fine-grained, data-driven memory control without retraining or architectural modifications. The source code and artifacts can be found in 
\href{https://github.com/Large-scale-Sustainable-Computing-LSC/ARKV}{GitHub}.

\end{abstract}


\section{Introduction}\label{s:intro}

Large Language Models (LLMs) have rapidly become the foundation of modern Natural Language Processing (NLP) and, increasingly, general-purpose artificial intelligence. Models such as \textit{LLaMA3}~\cite{touvron2023llamaopenefficientfoundation}, \textit{GPT-4}~\cite{openai2024gpt4technicalreport}, and \textit{Qwen3}~\cite{yang2025qwen3technicalreport} demonstrate impressive capabilities in generating coherent text, solving reasoning problems, writing code, and engaging in complex dialogue. At the same time, the emergence of deep-research agents and agentic-AI systems, in which LLMs autonomously plan, retrieve, and act across extended workflows and long documents, has significantly increased the demand for large context windows and persistent memory across sessions. For example, the Tongyi DeepResearch model \cite{tongyideepresearchteam2025tongyideepresearchtechnicalreport} is explicitly designed for deep information-seeking tasks with long-horizon workflows. Meanwhile, benchmarks such as AGENTIF \cite{qi2025agentifbenchmarkinginstructionfollowing} show that agentic scenarios often involve instructions with thousands of words and complex tool specifications. Such trends underscore that supporting ultra-long contexts and durable memory has become a central requirement in modern AI agent systems. Consequently, the need to support ultra-long contexts becomes a central challenge, one that dramatically boosts both computational and memory costs during inference. The feasibility of long-context inference is increasingly limited by KV cache memory, which grows linearly with sequence length and batch size and often dominates GPU memory consumption during deployment \cite{10.1145/3600006.3613165, 10.5555/3618408.3619696}.

To reduce this memory overhead, prior work has explored two main approaches. \textit{First}, \texttt{eviction-based sparsification}\cite{zhang2023h2oheavyhitteroracleefficient,guo2024attentionscoreneedtoken,huang2025locretenhancingevictionlongcontext}, which calculates the importance of tokens,  and retains only a subset of important tokens while discarding less important ones. They are built on the foundation that a subgroup of the KV cache is more important than others, could represent the attention distribution of all required tokens, and is sufficient for a successful generation. Although this eviction-based methods substantially reduce memory consumption, it risks losing critical contextual information, and precisely identifying the KV importance remains ultimately impossible, as the future of KV is undetermined. \textit{Second}, \texttt{Quantization-based} strategies \cite{liu2023kivi, hooper2025kvquant10millioncontext, zhang2024lorclowrankcompressionllms,yang2024tokenleftbehindreliable}, which compress KV tensors to smaller precision. These strategies preserve contextual information across all tokens, ensuring that long-range dependencies remain accessible to the model. However, prior studies have shown that aggressive low-bit quantization may distort attention distributions, due to outlier-dominated statistics, leading to degraded output generation quality, and inference instability \cite{bondarenko2023quantizabletransformersremovingoutliers, xiao2024smoothquantaccurateefficientposttraining, Shao2025WhenTT}. Beyond distribution distortion, its footprint and attention computation cost still grow linearly with sequence length. \textit{Third}, \texttt{Hybrid} approaches\cite{zhang2024q} combine sparsification and quantization, motivated by their complementary strengths in further reducing memory usage. However, existing hybrid methods typically rely on fixed heuristics or layer-agnostic policies, which limit their adaptability to heterogeneous importance and precision sensitivity across layers, heads, and decoding stages.

\noindent \textbf{Challenges:} We identify two core challenges that limit existing methods:
(i) Token importance is inherently layer-sensitive. The same token may contribute differently across transformer layers, yet current approaches typically apply uniform importance selection.
(ii) The quantization tolerance of tokens varies, and misaligning token precision with their actual importance can cause significant attention degradation. These issues call for a more fine-grained, adaptive approach to KV cache management.

\noindent \textbf{Contributions:} We propose Adaptive KV Cache under Budget (ARKV), a lightweight framework that dynamically assigns cache precision to tokens at runtime. During a short prefill phase, ARKV collects simple statistics (entropy, variance, kurtosis) from each attention layer to compute a layer-specific Original–Quantization (OQ) ratio, indicating how many tokens should be retained in full precision. During decoding, tokens are ranked via an efficient heavy-hitter score, and each is assigned to one of three states — Original, Quantization, or Eviction — under a strict memory budget. This tri-state, layer-aware design enables flexible adaptation without retraining or architectural changes.

Experiments on LLaMA3 and Qwen3 across long- and short-context tasks show that ARKV retains $\sim$97\% of baseline accuracy on LongBench, reduces KV memory usage by 4$\times$, and achieves up to 86\% of baseline throughput. On math reasoning (GSM8K), ARKV maintains high accuracy where uniform quantization fails. These results highlight the practical viability of ARKV for scalable and memory-efficient LLM deployment.

In summary, our key contributions are as follows:

\begin{itemize}
    \item We propose ARKV, a tri-state KV cache management framework that unifies eviction and quantization through dynamic token-level precision control.
    
    \item We introduce a lightweight, layer-aware Original–Quantization (OQ) ratio, derived from attention statistics, to guide per-layer budget allocation under memory constraints.

    \item We design a fast, online heavy-hitter scoring mechanism that ranks token importance and assigns each token to one of three states: Original, Quantization, or Eviction.

    \item We demonstrate that ARKV achieves $\sim$97\% accuracy retention with 4× memory savings on long-context tasks, while maintaining $\sim$14.4\% quantization ratio, near-baseline throughput, and robustness across diverse workloads. 
\end{itemize}

\section{Background and Motivation}\label{s:background}

\subsection{LLMs and KV Cache}

Transformer-based LLMs generate text autoregressively by repeatedly applying multi-head attention over past tokens. To avoid recomputing attention over the entire prefix at each decoding step, models store the key and value vectors of previously generated tokens in a dedicated KV cache. This cache allows constant-time lookups during decoding, significantly boosting throughput.

However, the KV cache size grows linearly with both context length and model size. For a decoder with $L$ layers, hidden dimension $d$, and $H$ attention heads, storing $T$ past tokens in bfloat16, batching $B$ requests requires $\mathcal{O}(2 \cdot B \cdot L \cdot H \cdot T \cdot d)$ bytes. For instance, the Llama-3.1 70B model, with an input batch size of 128 and a sequence length of 1024, results in 40GB of KV cache. As sequence lengths grow into tens or hundreds of thousands of tokens, especially in agentic workflows, retrieval-augmented generation, or multi-document QA. The KV cache can consume substantial memory, easily exceeding the capacity of a single GPU. This makes long-context inference a memory-bound problem: while model parameters are fixed at deployment, the per-input KV cache can scale unbounded, limiting throughput, batching, and deployability on resource-constrained hardware.

To address this challenge, prior work has explored several strategies to reduce the KV memory load. \texttt{Eviction-based} methods identify and remove less important past tokens from the cache. These techniques leverage heuristics like attention weights or recency to retain ``heavy-hitter" tokens while discarding others. While simple and fast, they risk removing tokens that may later become relevant. \texttt{Quantization-based} methods compress all or parts of the cache to lower-bit formats. This preserves all tokens, but aggressive quantization often introduces numerical instability, especially in sensitive reasoning tasks. \texttt{Hybrid} strategies combine eviction and quantization, selectively applying low precision to less important tokens while keeping others at full precision. These methods show promising trade-offs but often rely on static scoring rules or uniform layer treatment.

Despite their progress, these techniques are often limited by one or more of the following: 1) Static heuristics that cannot adapt to task, input, or layer variation. 2) Uniform treatment of layers or tokens, ignoring heterogeneous sensitivity. 3) Lack of online adaptation during generation. 4) Overhead from auxiliary models or tuning procedures, which adds complexity.

\subsection{Motivation}

Prior work has shown that both token selection and low-precision storage are effective in isolation, few methods provide a unified, adaptive mechanism that dynamically balances both under a global memory budget. Existing approaches often statically predefine importance metrics or compression ratios, failing to capture the token-level and layer-level heterogeneity in modern transformers.

These challenges motivate our research questions: 1) Can we design a unified tri-state caching scheme (retain, quantize, evict) that adjusts dynamically per token and per layer? 2) How can we infer the relative sensitivity of each layer to cache compression, in a lightweight, model-agnostic way? 3) Can we achieve strong long-context reasoning performance while staying within a strict memory budget, without retraining or sacrificing throughput?

To answer these questions, we propose ARKV, a lightweight framework that dynamically adjusts KV cache precision during decoding. ARKV leverages low-cost attention statistics gathered during a short prefill phase to allocate layer-wise precision budgets. During decoding, it ranks tokens using fast online importance scoring and assigns each to one of three states: Original, Quantization, or Eviction. This tri-state mechanism enables fine-grained, data-driven precision control, achieving strong accuracy–efficiency trade-offs across diverse tasks.

\section{Related Work}\label{s:related}
Existing research on KV cache optimization typically follows either eviction-based or quantization-based strategies. While some recent studies have investigated hybrid methods, these often rely on static configurations, making them impractical for dynamic contexts. In this section, we provide a detailed discussion of these related works.

\textbf{Eviction-based methods}. These methods reduce memory usage by selectively removing tokens from the KV cache, under the assumption that not all past tokens contribute equally to future predictions. H$_2$O \cite{zhang2023h2oheavyhitteroracleefficient} introduced the concept of heavy-hitter tokens, identifying and retaining only those with high accumulated attention scores, thus preserving around 20\% of the cache while maintaining performance without requiring retraining. Locret \cite{huang2025locretenhancingevictionlongcontext} further improves on this idea by finetuning LLMs with retaining heads that are dedicated modules trained to predict token importance. It enables highly selective and learned eviction policies, though at the cost of model modification. SnapKV \cite{li2024snapkvllmknowslooking} preemptively selects important tokens based on prompt-level attention analysis before generation begins. While lightweight, it applies a static keep-or-drop decision that cannot adapt to new attention patterns during generation. PyramidKV \cite{yang2024pyramidinferpyramidkvcache} proposes a layer-wise eviction strategy that allocates non-uniform cache budgets across transformer layers based on attention consolidation patterns, achieving strong compression without accuracy loss. However, these methods share common limitations: they often discard potentially useful tokens that fall below selection thresholds, lack quantization or compression mechanisms, and are generally unable to adjust dynamically during inference.

\textbf{Quantization-based methods}. These address memory overhead by compressing all cached key and value tensors to lower-bit representations. KVQuant \cite{hooper2025kvquant10millioncontext} applies token-level quantization down to 3 bits using learned scales and grouped quantization schemes, enabling extremely long context inference on single GPUs with minimal perplexity increase. KIVI \cite{liu2023kivi} introduces a 2-bit compression format with separate schemes for keys and values, achieving aggressive memory reduction without requiring retraining. However, both methods uniformly compress all tokens, including those critical to generation, which can introduce significant numerical degradation in reasoning-heavy tasks. Moreover, since token count remains unchanged, attention computation cost still grows linearly with context length. MiKV \cite{yang2024tokenleftbehindreliable} mitigates some of these issues by adopting a mixed-precision cache, where important tokens are kept in high precision and less important ones are reduced to low precision instead of being dropped. This preserves coherence better than hard eviction, but lacks token pruning altogether, meaning cache size still scales with input length. Furthermore, its fixed two-level precision scheme cannot capture the full spectrum of token sensitivities.

\textbf{Hybrid strategies}. These aim to combine the benefits of selective eviction and low-bit compression. Q-Hitter \cite{zhang2024q} introduces a sparse caching scheme that identifies ``attention sinks", tokens that are both frequently attended and not safe to quantize, and retains them at full precision. All other tokens are either evicted or quantized to 4-bit format, achieving up to 20× memory savings with minimal accuracy loss. However, Q-Hitter relies on predefined per-layer heuristics for quantization friendliness and attention accumulation, which are set offline and remain mostly static during inference. This limits its ability to respond to shifts in attention focus, potentially leading to suboptimal token state assignment. It also requires maintaining sensitivity maps and layer-specific scores, thereby introducing complexity and coupling with the model structure.

In contrast, our ARKV provides a unified, lightweight, and online tri-state caching mechanism. It dynamically assigns each cached token to one of three states based on real-time attention statistics and token importance scores during decoding. Crucially, ARKV is layer-aware: it computes attention-based entropy, variance, and kurtosis per layer during a brief prefill phase to derive per-layer OQ ratios, guiding how precision budgets are distributed. This enables fine-grained, adaptive control over cache memory allocation, allowing ARKV to tailor compression decisions to both token-level importance and layer-level sensitivity without retraining or static rules. Empirically, ARKV outperforms fixed-strategy baselines across a range of tasks, achieving high memory efficiency with minimal loss in accuracy or throughput.

\section{Methodology}\label{s:methodology}

In this section, we present the design of ARKV, a tri-state cache management framework for LLM inference under memory constraints. We describe how it estimates per-layer compression sensitivity, ranks token importance online, and enforces budget-aware precision assignment during decoding.

\subsection{System Model}
\begin{figure}[ht]
    \centering
    \includegraphics[width=0.95\linewidth]{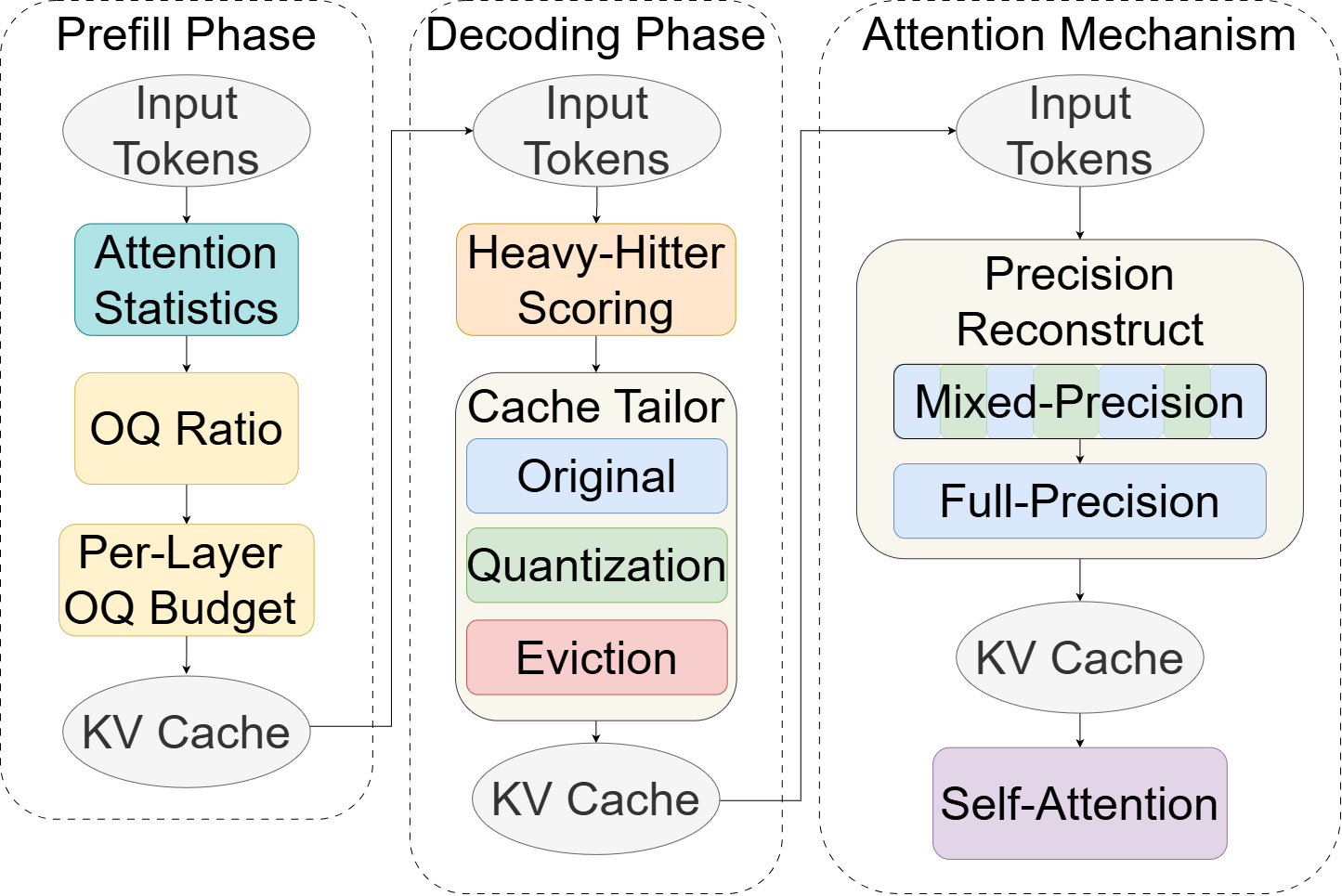}
    \caption{Overview of the ARKV framework. During the Prefill Phase, attention statistics are used to compute an Original–Quantization (OQ) ratio and allocate per-layer cache budgets. In the Decoding Phase, tokens are scored by importance and dynamically assigned to Original, Quantization, or Eviction states. The reconstructed KV cache feeds into the attention mechanism with mixed-precision handling to balance memory efficiency and model fidelity.}
    \label{fig:framework}
\end{figure}

Figure~\ref{fig:framework} presents a high-level overview of our ARKV system.  It consists of three key components: (1) \textit{Per-layer OQ ratio estimation} to determine each layer's compression sensitivity, (2) \textit{Token importance scoring} based on online attention statistics, and (3) \textit{Tri-state cache assignment} to fit a memory budget via selective precision control. As seen in Figure~\ref{fig:framework}, ARKV integrates into the decoder pipeline without model modification or retraining as it operates solely at the KV cache level and does not alter model parameters or attention computations. By reconstructing mixed-precision KV tensors that conform to the original cache format, ARKV remains fully compatible with standard self-attention kernels.

\subsection{Problem Formulation}
Let a Transformer model have $L$ layers and a global memory budget $B$ (in bytes). Each cached token can be in full precision (cost $C_{\text{orig}}$), quantized (cost $C_{\text{quant}}$), or evicted (cost 0). The objective is to minimize degradation in generation quality while ensuring the total cache size \( M_t \leq B \). For layer $\ell$, we denote $n_\ell^o$, $n_\ell^q$ as the number of original and quantized tokens.

\begin{equation}\label{eq:formulation}
    \sum_{\ell=1}^{L} \left(n^{\ell}_{o} C_{\text{orig}} + n^{\ell}_{q} C_{\text{quant}}\right) \le B.
\end{equation}

The equation \ref{eq:formulation} highlights that different layers and tokens together contribute to the total KV cache memory footprint and share a global memory budget. Our strategy, therefore, prioritizes both layer-level and token-level features.

\subsection{Attention Feature Calculation}
Existing works \cite{li2024snapkvllmknowslooking, yang2024pyramidinferpyramidkvcache} show that decoding patterns can be effectively captured by analyzing the attention patterns of recent queries, a short ``observation window". Hence, we focus on the window-restrained attention scores with the region $[Q-W, Q)$, and collect statistical features of each attention layer. Let $A^{(\ell)}\in \mathbb{R}^{H\times Q \times K}$ be the post-softmax attention tensor at layer $\ell$, with heads $H$ $\times$ queries $Q$ $\times$ keys $K$. Then we could denote the window-constrained attention by equation \ref{eq:windowed_att}

\begin{equation}\label{eq:windowed_att}
    \tilde{A}^{(\ell)} = \tilde{A}^{(\ell)}[:, Q-W, Q, 0:K-W]
\end{equation}

From $A^{(\ell)}$, we extract three statistical features (aggregated over heads and recent queries) for each key position $k\in [0, K-W)$ as follows. 

\begin{equation}\label{eq:entropy}
\begin{split}
    \mathcal{H} &= -\sum_{k=0}^{K-W-1}p_k\log p_k \\
    p_k&=\frac{1}{Z}\sum_{h,q}\tilde{A}_{h,q,k}^{(\ell)} \\
    Z&=\sum_{k=0}^{K-W-1}\sum_{h,q}\tilde{A}_{h,q,k}^{(\ell)}
\end{split}
\end{equation}

Entropy $\mathcal{H}$ \ref{eq:entropy} shows the spatial attention dispersion, where $Z$ is a normalization constant. 

\begin{equation}\label{eq:variance}
    \mathcal{V}^{(\ell)} = \frac{1}{K-W}\sum_{k=0}^{K-W-1}\left(p_k-\bar p\right)^2
\end{equation}

Variance $\mathcal{V}$ \ref{eq:variance} represents the difference of all token weights.

\begin{equation}\label{eq:kurtosis}
    \mathcal{K} = \frac{\mathbb{E}\left[(p_k-\bar p)^4\right]}{\left(\mathbb{E}\left[(p_k-\bar p)^2\right]\right)^2}
\end{equation}

Kurtosis $\mathcal{K}$ \ref{eq:kurtosis} gives a picture of attention's distribution. 
Together, entropy, variance, and kurtosis provide a compact characterization of attention concentration and dispersion, which we use as lightweight proxies for layer-wise sensitivity.

\subsection{Per-layer OQ Ratio Estimation}

Based on the statistical features, we calculate original-quantization scores to evaluate the $OQ$ ratio $\mathcal{P}$. $OQ$ ratio indicates the ratio of the three states in each layer. Since we protect the last $W$ recent tokens, $OQ$ scoring is only applied to the region $[0:K-W)$. Here, 

\begin{equation}\label{eq:oq_score}
    q_\ell = \left(\mathcal{H}^{(\ell)}\right)^{\frac{1}{\tau_1}}
    \left(\mathcal{V}^{(\ell)}\right)^{\frac{1}{\tau_2}}
    \left(\mathcal{K}^{(\ell)}\right)^{\frac{1}{\tau_3}}
\end{equation}

The $OQ$ score $q_\ell$ \ref{eq:oq_score} of a layer $\ell$ is the combination of entropy $\mathcal{H}^{(\ell)}$, variance $\mathcal{V}^{(\ell)}$, and kurtosis $\mathcal{K}^{(\ell)}$. Temperature parameters $\tau_1$, $\tau_2$, $\tau_3$ are used to balance the effect of the three statistical features. 

\begin{equation}\label{eq:oq_ratio}
    \rho_\ell = \frac{q_\ell}{\max_{k=\{0,1,\dots,L\}}(q_k)}
\end{equation}

The $OQ$ ratio $\rho_\ell$ \ref{eq:oq_ratio} of an attention layer indicates the ratio between original and quantized precision of tokens in the KV cache. The $OQ$ scores demonstrate ``how important" an attention layer is. The more important an attention layer is, the more tokens should be kept in their original precision. A lower $OQ$ score indicates more quantized tokens in the KV cache, vice versa. 

\begin{equation}\label{eq:budget}
    B_o^{(\ell)} = B\cdot\mathcal{\rho_\ell}, \quad
    B_q^{(\ell)} = B-B_o^{(\ell)}
\end{equation}

Equation \ref{eq:budget} shows that with a given KV cache budget $B$, which is the same for all layers, we allocate the budget $B_o$ for original precision tokens and the budget $B_q$ for quantized precision tokens.

\subsection{Token Importance via Heavy-Hitter Scoring}

Heavy-hitter scores are per-layer importance estimates derived from cumulative attention distributions that quantify how important each token is attended to across layers and timesteps. In ARKV, they serve as the primary criterion for assigning token state, thus guiding memory allocation under budget constraints.

\begin{equation}\label{eq:hh_score}
\begin{split}
    \mathcal{S}_k^{(\ell)}&=\mu_k^{(\ell)} + \gamma \cdot \left(\sigma_k^{(\ell)}\right)^2 \\
    \mu_k^{(\ell)} &= \mathbb{E}_{h,q}\tilde{A}_{k,q,k}^{(\ell)} \\ 
    \sigma_k^{(\ell)} &= \text{Var}_{h,q}\tilde{A}_{k,q,k}^{(\ell)}\\
\end{split}
\end{equation}

The $hh$ score $\mathcal{S}_k^{\ell}$ for an attention layer $\ell$ and token $k$ is defined with equation \ref{eq:hh_score}. It is the combination of the mean $\mu_k^{(\ell)}$ and variance $\sigma_k^{(\ell)}$ of a token $k$ and an attention layer $\ell$ over heads $h$ and queries $q$. $\gamma$ is a tunable hyperparameter. For grouped-query attention(GQA), scores are averaged within each group to preserve KV sharing.

\subsection{Tri-State Cache Tailor and Mixed-Precision Integration}

At each step, tokens outside a protected window of size $W$ are reassigned as follows:
\begin{itemize}
\item Sort eligible tokens in each layer by $S_\ell(k)$.
\item Assign top $OB_\ell$ tokens as Original, next $QB_\ell$ as Quantized.
\item Evict remaining tokens.
\end{itemize}
The values of $OB_\ell$ and $QB_\ell$ are computed to satisfy per-layer memory quotas derived from $\rho_\ell$ and the global budget $B$.
\begin{equation}\label{eq:keep}
\begin{split}
    \mathcal{I}_{all}^{\ell} &= \{0,1,\dots,K-W-1\} \\
    \mathcal{I}^{\ell} &= \text{Top-b}\left(\mathcal{I}_{all}^{\ell}, \mathcal{S}^{\ell}\right), \quad b=\lfloor \alpha (K-W) \rfloor \\
    \mathcal{I}_o^{\ell} &= \text{Top-}B_o^{(\ell)}\left(\mathcal{I}^{\ell}, \mathcal{S}^{\ell}\right) \\
    \mathcal{I}_q^{\ell} &= \mathcal{I}^{\ell} \setminus \mathcal{I}_o^{\ell} \\
    \mathcal{I}_q^{\ell} &= \mathcal{I}_{all}^{\ell} \setminus \mathcal{I}^{\ell} \\
    \mathcal{I}_o^{\ell} &= \mathcal{I}_o^{\ell} \cup [K-W:K)
\end{split}
\end{equation}

Equations \ref{eq:keep} demonstrate the process of the KV cache tailor, which will be triggered when the KV cache reaches the limit. Ranking by $\mathcal{S}_k^{(\ell)}$ of the tokens, we select the tokens that should be kept $\mathcal{I}^{\ell}$, kept as original precision $\mathcal{I}_o^{\ell}$, kept as quantized precision $\mathcal{I}_q^{\ell}$, and evicted $\mathcal{I}_e^{\ell}$. 
To prevent frequent tailoring, we reserve a small buffer for incoming new tokens using a slack factor $\alpha$. It is set to 0.75 in the experiments.

ARKV merges Original and Quantized tokens before each attention step. Quantized entries are dequantized and concatenated to form a contiguous cache. The attention mechanism is unmodified.

\subsection{Algorithm}

Algorithm \ref{alg:eq} demonstrates the end-to-end pipeline of ARKV with pseudocode. In general, it has three phases in total: 1) \textbf{Prefill phase}: Each layer’s Origin budget is determined by an $OQ$ score that reflects attention distributions feature. 2) \textbf{Decoding phase}: At each step, $hh$ scores are computed only on the evictable region $[0:K-W)$. The tri-state cache tailor applies here. 3) \textbf{Reconstruction Before Attention}: Quantized entries are dequantized on-the-fly using per-token scales and merged with Original entries in their original sequence order. This guarantees a logically contiguous KV cache for attention.

\begin{algorithm}[htbp]
\caption{ARKV Framework Algorithm}
\label{alg:eq}
\KwIn{KV cache length $K$, window size $W$, total budget $B$, attention scores $A$}
\KwOut{Unified KV cache $\tilde{K}, \tilde{V}$}

\BlankLine
\textbf{Prefill Phase: $OQ$ Ratio Allocation}\\
\For{each layer $\ell=1,\dots,L$}{
    Compute entropy $\mathcal{H}^{(\ell)}$, variance $\mathcal{V}^{(\ell)}$, kurtosis $\mathcal{K}^{(\ell)}$ from $A$\;
    $q_\ell \gets (\mathcal{H}^{(\ell)})^{1/\tau_1} (\mathcal{V}^{(\ell)})^{1/\tau_2} (\mathcal{K}^{(\ell)})^{1/\tau_3}$\;
}
$OQ$ Ratios: $\rho_\ell = q_\ell / {\max_{k=\{0,1,\dots,L\}}(q_k)}$\;
Assign origin budget: $B_o^{(\ell)} = \rho_\ell \cdot (B-W)$\;

\BlankLine
\textbf{Decoding Phase: KV Cache Tailor}\\
For current step with cache length $K$\;
Compute heavy-hitter score for window-excluded tokens $[0:K-W)$: \\
\quad $\mathcal{S}_k = \mu_k + \gamma \cdot \sigma_k^2$ (averaged across heads, smoothed, GQA-aligned)\;
Let: $\mathcal{I}_{all} = \{0,1,\dots,K-W-1\}, \quad b=\lfloor \alpha (K-W) \rfloor$
Select keep set: $\mathcal{I} = \operatorname{Top}\text{-}b(\mathcal{I}_{all},S)$\;
Split: \\
\quad $\mathcal{I}_{o} = \operatorname{Top}\text{-}B_o^{(\ell)}(\mathcal{I}, S)$\; 
\quad $\mathcal{I}_{q} = \mathcal{I} \setminus \mathcal{I}_{o}$\; 
\quad $\mathcal{I}_{e} = \mathcal{I}_{all} \setminus \mathcal{I}$\;
Always protect window: $\mathcal{I}_{o} \gets \mathcal{I}_{o} \cup [K-W:K)$\;

\BlankLine
\textbf{Reconstruction Before Attention}\\
\For{$k \in \mathcal{I}_{q}$}{
    Dequantize: $\tilde{x}_{k} = \hat{q}_k \cdot s_k$ (per-token scale)\;
}
Scatter $\mathcal{I}_{o}$ and $\mathcal{I}_{q}$ into unified sequence order\;
Return reconstructed $\tilde{K}, \tilde{V}$ for attention computation\;

\end{algorithm}

With this algorithm in place, our system ensures at each step that the KV cache is trimmed and compressed to respect the memory limit, while aiming to keep the most relevant information in the highest fidelity.

\section{Performance Evaluation}\label{s:experiments}

\subsection{Implementation}

We implemented ARKV on top of PyTorch \cite{10.5555/3454287.3455008}, using the Hugging Face framework \cite{huggingfaceacc} for its simplicity. It is designed as a drop-in KV cache manager for decoder-only LLMs. We integrate ARKV by conforming to the standard \texttt{Cache}/\texttt{DynamicCache} interface and override the attention forward pass to accept a custom \texttt{past\_key\_values} object, while fully preserving the \texttt{generate()} API and existing attention kernels. Models are hosted locally using HuggingFace checkpoints in bfloat16, and all experiments are executed using standard autoregressive decoding. Evaluation workloads are generated from LongBench\cite{bai2024longbenchbilingualmultitaskbenchmark}, and lm-eval \cite{eval-harness} with deterministic decoding, and fine-tuning with Optuna\cite{optuna_2019}. The source code is available here: \textbf{\url{https://github.com/Large-scale-Sustainable-Computing-LSC/ARKV}.}

\subsection{Experimental Setup}
\noindent \textbf{Models.} We evaluate ARKV on two families of models: LLaMA3 \cite{grattafiori2024llama3herdmodels} (3.2-3B-Instruct, 3.1-8B-Instruct) and Qwen3 \cite{yang2025qwen3technicalreport} (Qwen3-4B-Instruct-2507, 3-8B). These models are representative of modern LLMs and strong baseline performers on the chosen tasks.
\newline
\noindent \textbf{Datasets.}
We sample reasoning and knowledge tasks on GSM8K \cite{cobbe2021trainingverifierssolvemath}, MMLU \cite{hendrycks2021measuringmassivemultitasklanguage}, and CommonsenseQA \cite{talmor2019commonsenseqaquestionansweringchallenge} with lm-eval \cite{eval-harness}. Also, we evaluate the long-context understanding of our approach on LongBench \cite{bai2024longbenchbilingualmultitaskbenchmark}, which contains a wide range of tasks with long context, including single-document QA, multi-document QA, summarization, few-shot learning, synthetic, and code. 
\newline
\noindent \textbf{Testbed.} The experiments are tested on DAS6 \cite{7469992}, with an NVIDIA A6000 48GB GPU and an AMD EPYC 7402P 24-Core Processor.
\subsection{Baselines}

In order to fully compare the performance of our system, we consider multiple strategies as baselines, as shown as follows:

\begin{itemize}
    \item \textbf{Base\_model} (No compression, abbreviated as `base'): Without cache limitation and all tokens are in $bfloat16$ form. This is the full model without any modifications.
    
    \item \textbf{Base\_origin }(origin-only heavy hitters, abbreviated as `origin'): Has cache limitation. All tokens are in $bfloat16$ form.
    \item \textbf{Base\_quant }(quantized-only heavy hitters, abbreviated as `quant'): Has cache limitation. Except for the window-protected region, all tokens are in $fp8$ form.
\end{itemize}
Our approach is marked as ARKV in the experiments.

\subsection{Evaluation Metrics and Benchmarks}
LLMs are usually measured by efficiency and accuracy metrics from multiple perspectives as summarized in Table~\ref{tab:metrics_benchmarks_intro}:
\begin{table}[h]
    \centering
    \begin{tabular}{l|l}
        \toprule
        \textbf{Metric / Benchmark} & \textbf{Description} \\
        \hline
        TPS (Tokens Per Second)  & Token generation rate \\
        GSM8k \cite{cobbe2021trainingverifierssolvemath} & Arithmetic reasoning \\
        MMLU \cite{hendrycks2021measuringmassivemultitasklanguage} & General knowledge\\
        CommonsenseQA \cite{talmor2019commonsenseqaquestionansweringchallenge}  & Commonsense questions\\
        LongBench \cite{bai2024longbenchbilingualmultitaskbenchmark} & Long-context understanding \\
        \bottomrule
    \end{tabular}
    \caption{Evaluation Metrics and Benchmarks}
    \label{tab:metrics_benchmarks_intro}
\end{table}

The primary efficiency metric is Tokens Per Second (TPS), which measures the average token generation rate during inference and reflects the model's runtime performance across different KV cache configurations. To assess task accuracy, four widely recognized benchmarks could work. GSM8k focuses on arithmetic reasoning problems, testing a model’s multi-step problem-solving capabilities. MMLU evaluates general knowledge across multiple academic and professional domains. CommonsenseQA challenges models with multiple-choice questions requiring everyday reasoning. Finally, LongBench provides a suite of tasks explicitly designed for evaluating long-context understanding, including retrieval QA, summarization, and multi-document comprehension. Together, these metrics provide a comprehensive view of both the functional accuracy and computational efficiency of the proposed ARKV framework.

\par In addition, we also perform memory measurements and relative performance. 
\begin{itemize}
    \item \textbf{Memory usage of KV cache:} How much of the KV cache is used in each state, expressed as the percentage of tokens evicted or as a percentage of the compressed cache. This reflects the memory footprint.
    \item \textbf{Relative performance:} The model’s accuracy on each task is scored in terms of the percentage of baseline performance retained. This allows a fair comparison of the performance retained under each memory-saving method.
\end{itemize}

Together, these metrics capture important characteristics that reflect the central tension in KV cache research. We aim to maintain high accuracy while significantly reducing memory usage and maintaining high throughput.

\subsection{Main Results}

We tested ARKV and the baselines with cache limits of 512, 1024, and 2048 tokens. The window size is set to 32 for our approach and all other baselines that require window size settings. After fine-tuning, models with $\tau_1=7.774$, $\tau_2=5.407$, $\tau_3=5.528$, and $\gamma=263.81$ shows the best performance.

\subsection{Performance on LongBench}
\begin{figure}[htbp]
    \centering
    \includegraphics[width=1\linewidth]{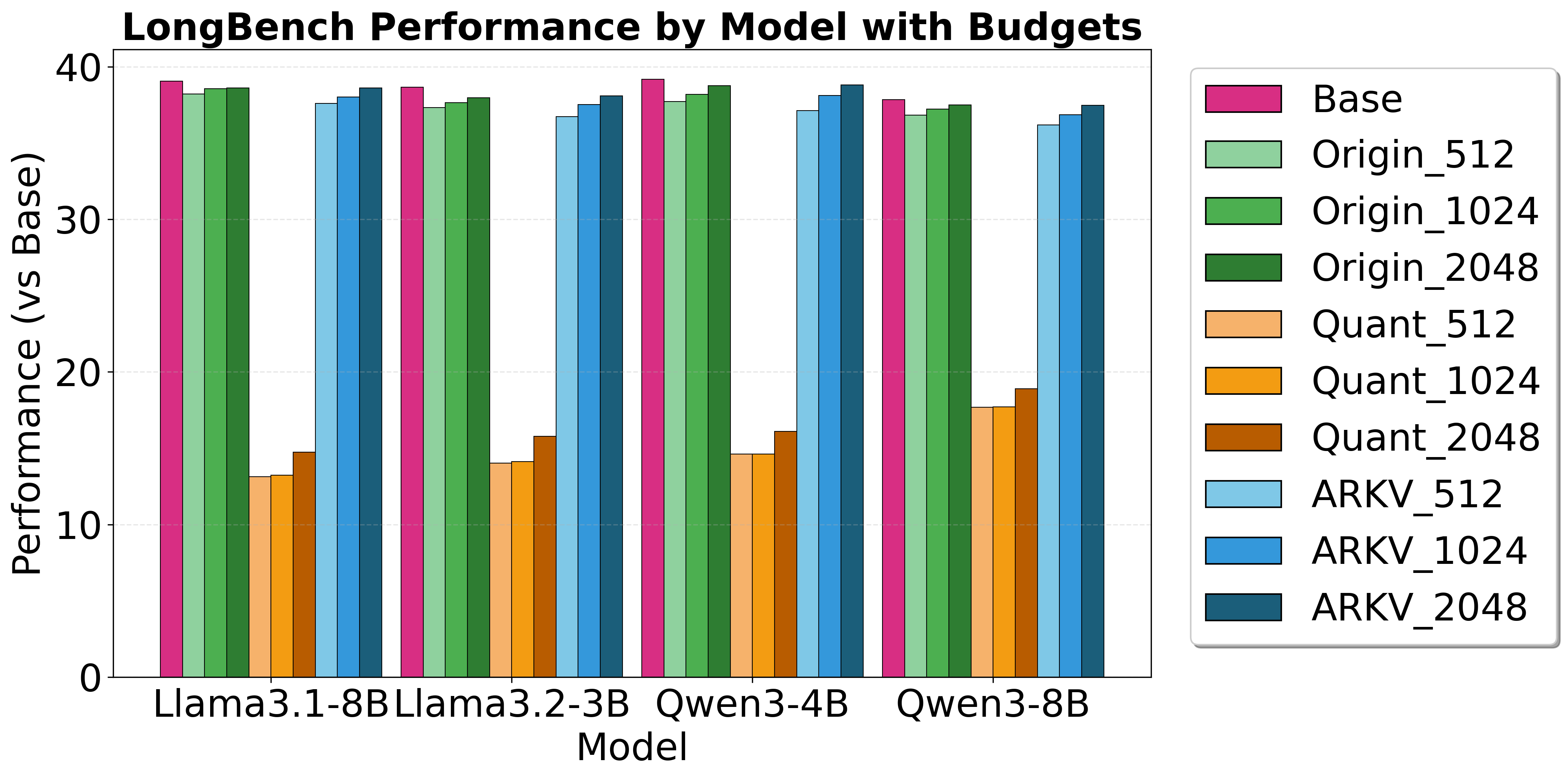}
    \includegraphics[width=1\linewidth]{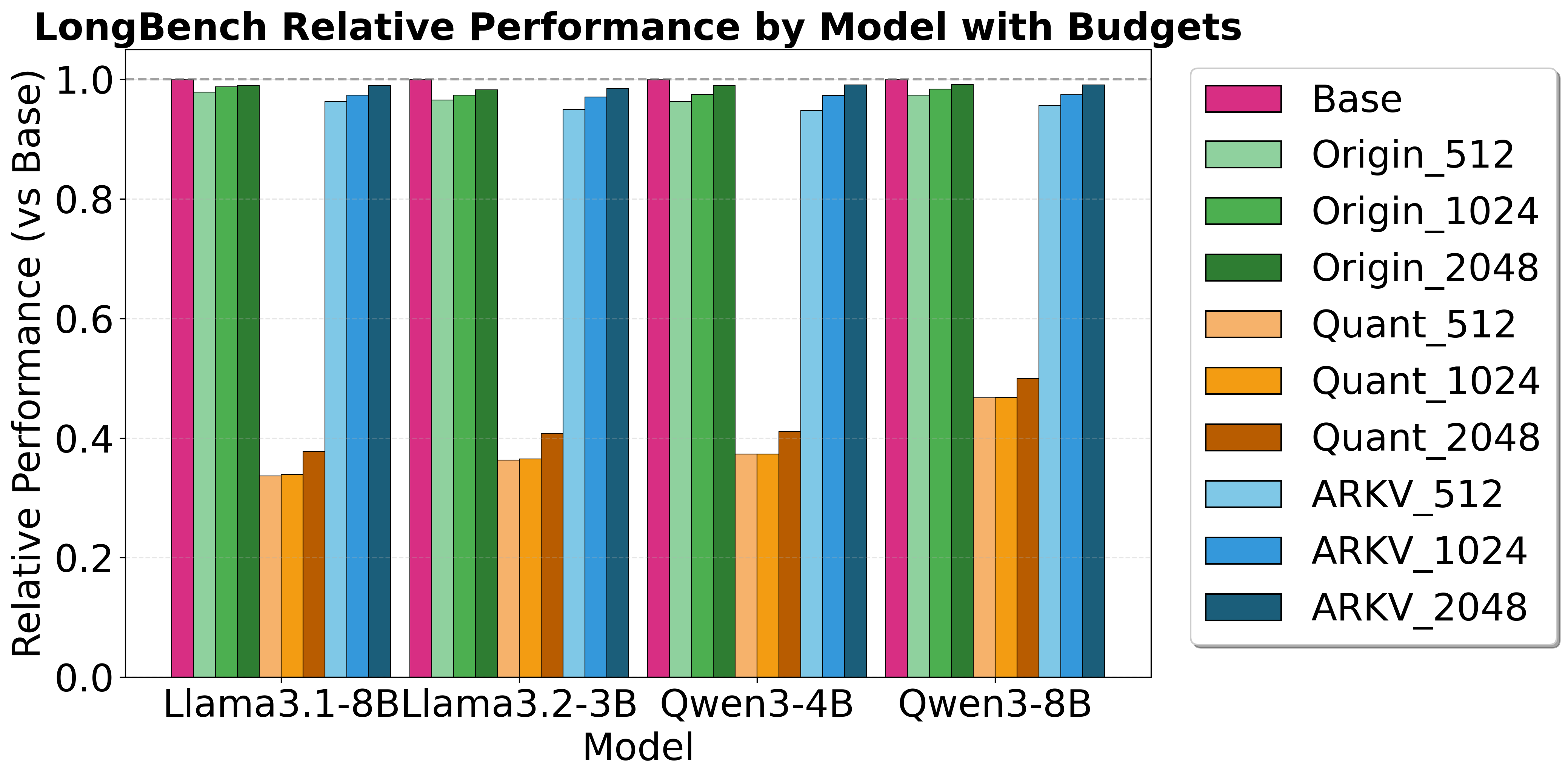}
    \caption{LongBench performance across different models and KV cache strategies. The top panel shows absolute scores, while the bottom panel presents performance relative to the full-cache Base model (normalized to 1.0). Each group includes results for Origin (green), Quant (orange), and ARKV (blue), evaluated at token budgets of 512, 1024, and 2048 (from light to dark). ARKV consistently outperforms quantization-only baselines and closely matches origin-only performance, even under tight memory budgets.}
    \label{fig:longbench_avg}
\end{figure}
LongBench is a challenging benchmark designed to assess the ability of LLMs to handle long-context problems across real-world multitasks. 
Figure~\ref{fig:longbench_avg} shows the average relative perform by cache budgets of all models. The results are consist cross different model families and parameter sizes. The experiment shows that ARKV performed relatively close to the base\_model, matching the base\_origin, and outperformed the base\_quant.

Table~\ref{tab:overall_rela_perf_longbench} shows that ARKV achieves performance very close to the base\_origin baseline while substantially outperforming base\_quant under limited cache budgets. Averaged across models, ARKV attains a relative score of 0.972 compared to 0.979 for base\_origin, whereas base\_quant drops sharply to 0.398. This indicates that the primary source of performance degradation is the reduced cache budget, rather than the selective mixed-precision mechanism. In particular, ARKV largely preserves the accuracy of the full-precision baseline, while uniformly quantizing the entire KV cache leads to severe performance collapse, confirming that indiscriminate quantization is unsuitable for long-context inference.

A finer-grained analysis by task type and language further highlights heterogeneous sensitivity to cache constraints. As shown in Table~\ref{tab:task_type_performance_summary}, tasks such as code generation and few-shot learning are relatively robust to limited budgets, whereas multi-document QA, single-document QA, and summarization exhibit more noticeable degradation. In these comparable challenging tasks, ARKV consistently retains most of its performance with a relative score of 0.93–0.95, whereas base\_quant suffers substantial drops with a relative score of 0.27–0.34. Table~\ref{tab:language_performance_summary} further reveals that Chinese tasks are the most affected by cache constraints: under a 1024-token budget, ARKV achieves 0.94 compared to the full-cache baseline of 1.00, whereas base\_quant degrades sharply to 0.28. English and programming-language tasks are less sensitive, where ARKV $\sim$0.98 and base\_quant $\sim$0.49 for English and $\sim$0.38 for code. It indicates that selective precision control significantly mitigates performance loss across diverse task distributions.

\begin{table}[htbp]
    \centering
    \begin{tabular}{c c c | c}
        \toprule
        \textbf{Base} & \textbf{Origin} & \textbf{Quant} & ARKV\\
        \hline
        1.00 & 0.979 & 0.398 & 0.972\\
        \bottomrule
    \end{tabular}
    \caption{Overall Relative Performance on LongBench}
    \label{tab:overall_rela_perf_longbench}
\end{table}

\begin{table}[htbp]
\centering

\begin{tabular}{l|c c c|c}
\toprule
\textbf{task type} & \textbf{Base} & \textbf{Origin} & \textbf{Quant} & \textbf{ARKV} \\
\hline
Code & 1.00 & 1.00 & 0.34 & 0.98 \\
Few shot & 1.00 & 1.00 & 0.48 & 1.00 \\
Multi-doc QA & 1.00 & 0.94 & 0.52 & 0.95 \\
Single-doc QA & 1.00 & 0.96 & 0.34 & 0.94 \\
Summarization & 1.00 & 0.95 & 0.27 & 0.93 \\
Synthetic & 1.00 & 1.04 & 0.59 & 1.03 \\
\bottomrule
\end{tabular}

\caption{Average relative performance by task type}
\label{tab:task_type_performance_summary}
\end{table}

\begin{table}[htbp]
\centering
\begin{tabular}{l|c c c|c}
\toprule
\textbf{language} & \textbf{Base} & \textbf{Origin} & \textbf{Quant} & \textbf{ARKV} \\
\hline
EN & 1.00  & 0.98 & 0.49 & 0.98 \\
Python/C\#/Java & 1.00  & 1.00 & 0.38 & 0.99 \\
Python/Java & 1.00  & 0.99 & 0.31 & 0.97 \\
ZH & 1.00 & 0.95 & 0.28 & 0.94 \\
\bottomrule
\end{tabular}
\caption{Average relative performance by language}
\label{tab:language_performance_summary}
\end{table}

\subsection{Performance on Short-Context Tasks.}

The Short-Context tasks include GSM8k \cite{cobbe2021trainingverifierssolvemath}, MMLU \cite{hendrycks2021measuringmassivemultitasklanguage}, and CommonsenseQA \cite{talmor2019commonsenseqaquestionansweringchallenge}. According to the experiment, 
the results on GSM8K are the only ones that are affected by the budgets and quantizations. This is because task lengths on MMLU and CommonsenseQA are shorter than the shortest budget limits, 512. The question lengths in MMLU range from 4 to 4.67K, with 84.3\% of the questions shorter than 471, whereas in CommonsenseQA they range from 24 to 321. In these cases, ARKV, base\_origin, and base\_quant behave the same as the base\_model, as these tasks only require understanding relatively short questions and retrieving knowledge. The model doesn't generate extremely long reasoning that would overflow the cache. 

\begin{figure}[htbp]
    \centering
    \includegraphics[width=1\linewidth]{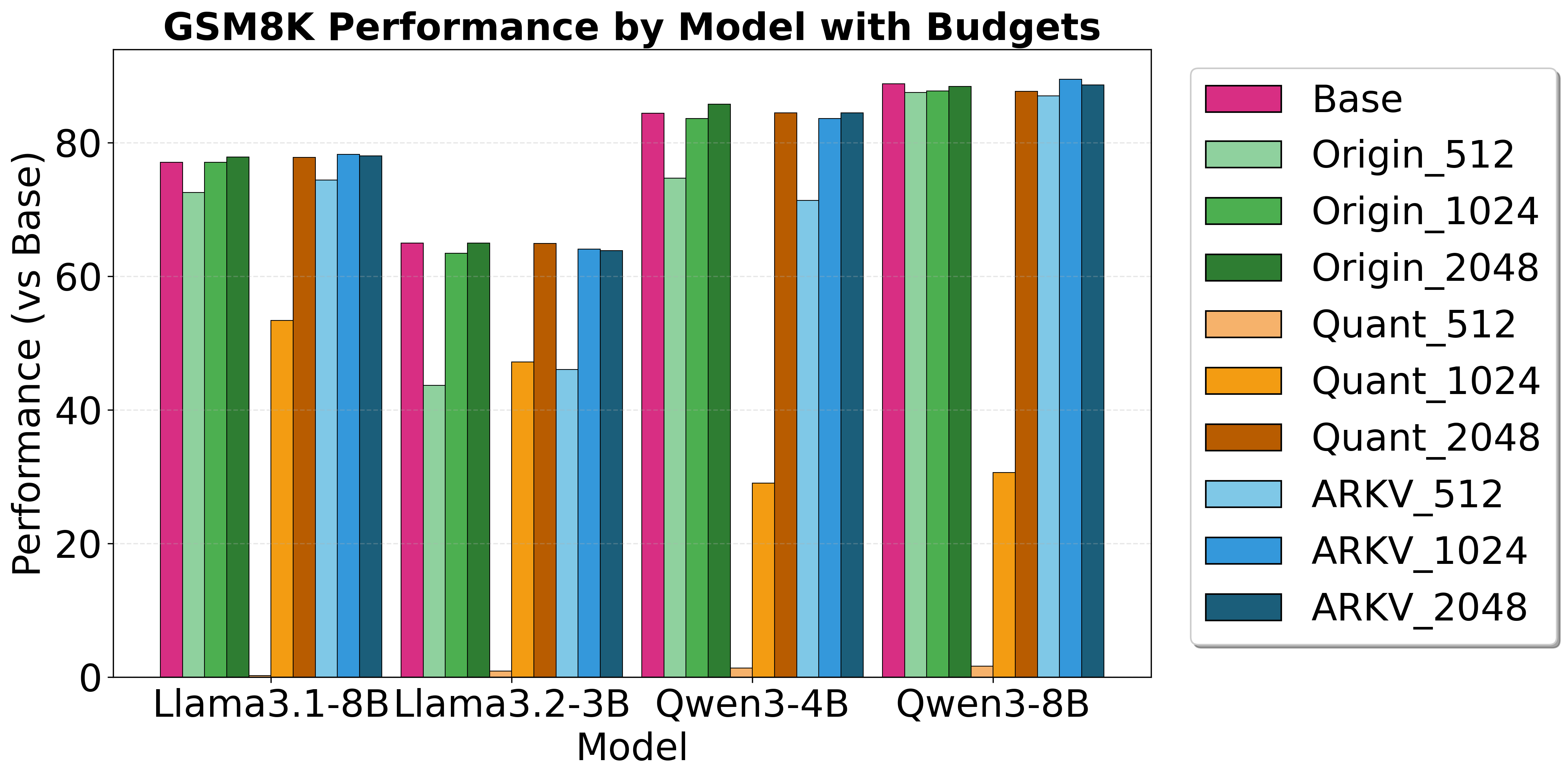}
    \includegraphics[width=1\linewidth]{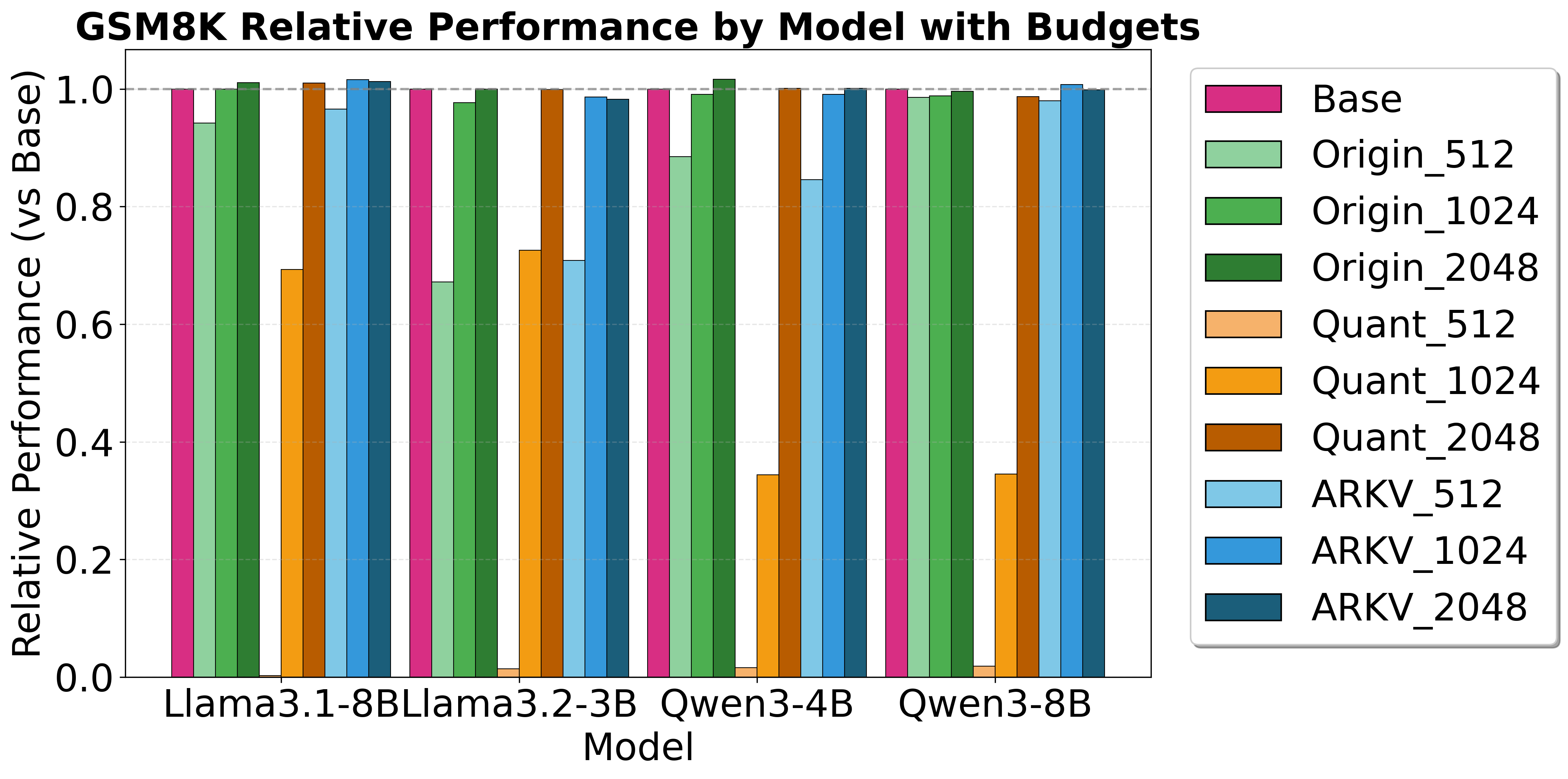}
    \caption{GSM8K performance under various KV cache strategies and budgets across different models. The top panel shows absolute accuracy scores, while the bottom panel presents performance relative to the full-cache Base model (normalized to 1.0). Green bars represent origin-only retention, orange bars indicate uniform quantization, and blue bars denote ARKV. Each budget level (512, 1024, 2048 tokens) is visualized from light to dark. ARKV consistently preserves high accuracy across all models, whereas quantization-only methods degrade sharply, especially for small budgets and smaller models.}
    \label{fig:ov_gsm8k_perf}
\end{figure}

Figure~\ref{fig:ov_gsm8k_perf} gives an overview of the overall performance on GSM8K. Table~\ref{tab:gsm8k_stati} provides the details of the performance. 
At the 512 budget, base\_quant’s performance was almost zero. Quantizing everything prevented the system from solving the mathematical calculations, possibly because tiny quantization errors in intermediate calculations compound to produce incorrect answers. As the budget increases to 2048, which exceeds the task lengths, all strategies become identical. In summary, GSM8K is sensitive to both budget size and quantization, and relies more on token precision. 
ARKV performed as well as base\_origin, underscoring that our adaptive approach preserved GSM8K performance nearly fully once the budget was moderately large, whereas quantizing everything severely degraded accuracy.

In summary, most tasks were robust to our method, since context wasn’t too long, but GSM8K highlighted that when reasoning and numeric precision are involved, quantizing everything is disastrous, while our approach is nearly as good as no quantization.

\begin{table}[ht]
\centering
\begin{tabular}{c|c c c|c}
\toprule
\textbf{Budget} & \textbf{Base} & \textbf{Origin} & \textbf{Quant} & \textbf{ARKV} \\
\hline
512  & \textbf{0.791} & 0.696 & 0.010 & \underline{0.697} \\
1024 & \textbf{0.791} & 0.780 & 0.401 & \underline{0.789} \\
2048 & \textbf{0.791} & \underline{0.794} & 0.789 & 0.788 \\
\bottomrule
\end{tabular}
\caption{Mean accuracy on GSM8K under different KV cache budgets. ARKV consistently matches Origin and substantially outperforms Quant, especially under tight memory constraints. The best result within each budget block is underlined.}
\label{tab:gsm8k_stati}
\end{table}

\subsection{TPS and Memory Savings.}

In addition to the performance over a wide range of tasks, we evaluate the efficiency of inference and memory savings. The token generated per second (TPS) is used to measure the speed of inference. As for memory savings evaluation, we calculate \textit{Quant Ratio}, calculated with $token\_quant/cache\_budget$, then compared with base\_origin, since $fp8$ occupied half the memory of $bfloat16$.

\begin{figure}[htbp]
    \centering
    \includegraphics[width=1\linewidth]{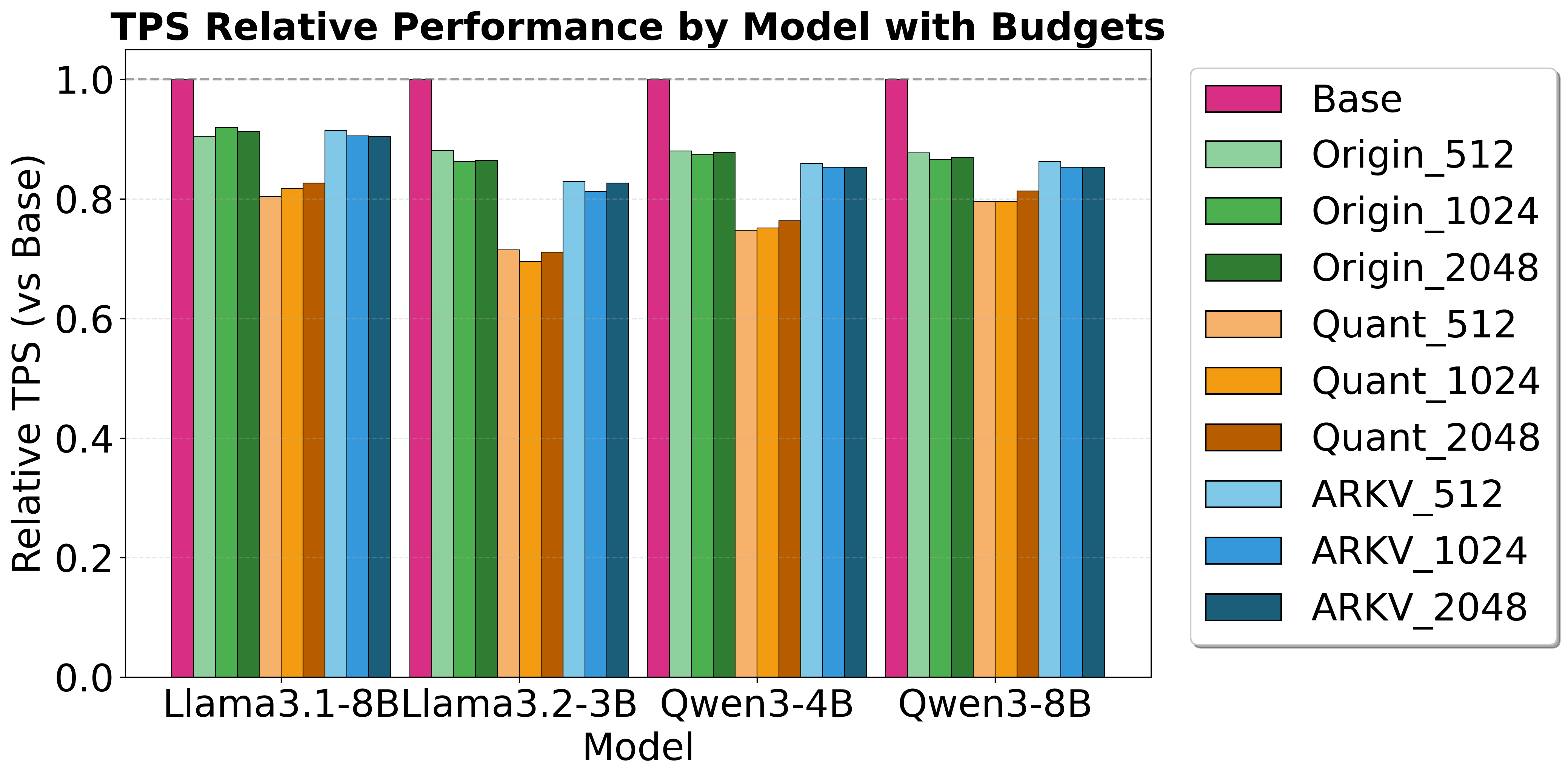}
    \caption{Relative throughput (TPS) across different models under various KV cache strategies and budgets. Each group shows performance of Base, Origin, Quant, and ARKV at budgets of 512, 1024, and 2048 tokens (from light to dark shades). Green bars represent full-precision eviction (Origin), orange bars show uniform quantization (Quant), and blue bars indicate ARKV. ARKV consistently retains high throughput ($\sim$85–88\%) across models while enforcing cache constraints.}
    \label{fig:ov_tps}
\end{figure}

\textbf{TPS}. Figure~\ref{fig:ov_tps} gives an overview of the TPS over all tasks. Table~\ref{tab:tps_performance_relative_to_base_method} provides the details of the TPS of all models, budgets, and strategies. As a result, our approach has 86.1\% relative TPS with 13.9\% throughput reduction, base\_origin has 88.2\% relative TPS with 11.8\% throughput reduction, and base\_quant has 77.0\% relative TPS with 23.0\% throughput reduction. ARKV has approximately the same TPS as base\_origin, with negligible degradation, while base\_quant experiences significant degradation due to the consumption of quant-dequant calculations. Moreover, our method's TPS is almost the same as pure eviction, meaning we haven't introduced a large overhead in exchange for memory savings. The degradation of performance on base\_origin, a pure eviction strategy, might be caused by the implementation. Since other pure eviction works show an improvement in throughput.

\begin{table}[htbp]
\centering

\begin{tabular}{l c|c c c|c}
\toprule
\textbf{Model} & \textbf{Budget} & \textbf{Base} & \textbf{Origin} & \textbf{Quant} & \textbf{ARKV} \\
\hline
LLaMA3.2-3B & 512  & 1.000 & 0.898 & 0.717 & 0.835 \\
LLaMA3.2-3B & 1024 & 1.000 & 0.864 & 0.683 & 0.810 \\
LLaMA3.2-3B & 2048 & 1.000 & 0.876 & 0.685 & 0.823 \\
\hline
LLaMA3.1-8B & 512  & 1.000 & 0.860 & 0.760 & 0.885 \\
LLaMA3.1-8B & 1024 & 1.000 & 0.890 & 0.773 & 0.883 \\
LLaMA3.1-8B & 2048 & 1.000 & 0.875 & 0.769 & 0.881 \\
\hline
Qwen3-4B   & 512  & 1.000 & 0.880 & 0.709 & 0.862 \\
Qwen3-4B   & 1024 & 1.000 & 0.875 & 0.708 & 0.849 \\
Qwen3-4B   & 2048 & 1.000 & 0.881 & 0.712 & 0.858 \\
\hline
Qwen3-8B   & 512  & 1.000 & 0.856 & 0.773 & 0.855 \\
Qwen3-8B   & 1024 & 1.000 & 0.842 & 0.765 & 0.846 \\
Qwen3-8B   & 2048 & 1.000 & 0.852 & 0.774 & 0.846 \\
\bottomrule
\end{tabular}
\caption{Relative inference throughput (TPS) of each method compared to the base model (1.0), across four LLM variants. ARKV maintains $\sim$85\% of base throughput on average, with significantly better efficiency than Quant under all settings.}
\label{tab:tps_performance_relative_to_base_method}
\end{table}

\begin{figure}[htbp]
    \centering
    \includegraphics[width=1\linewidth]{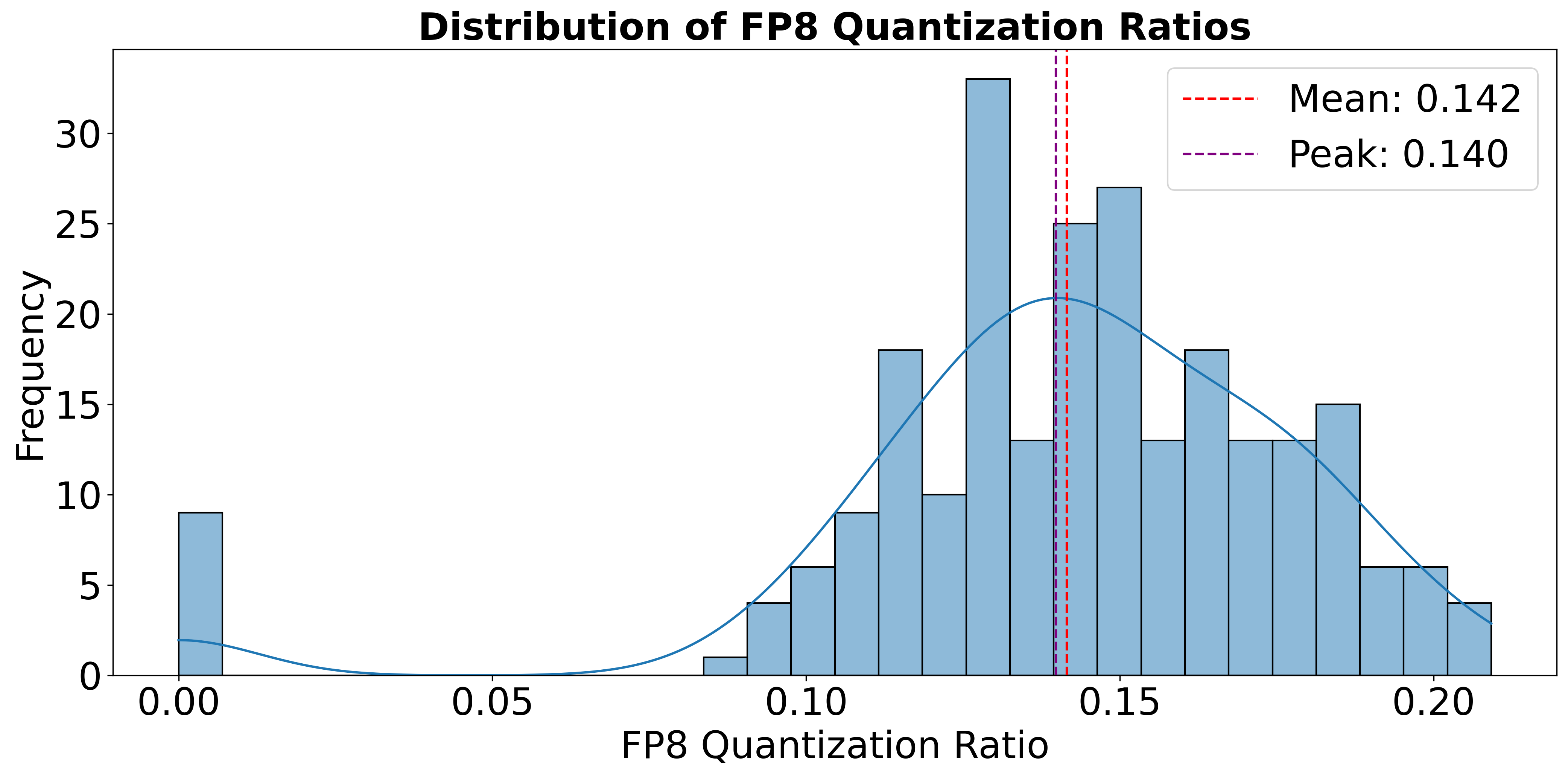}
    \includegraphics[width=1\linewidth]{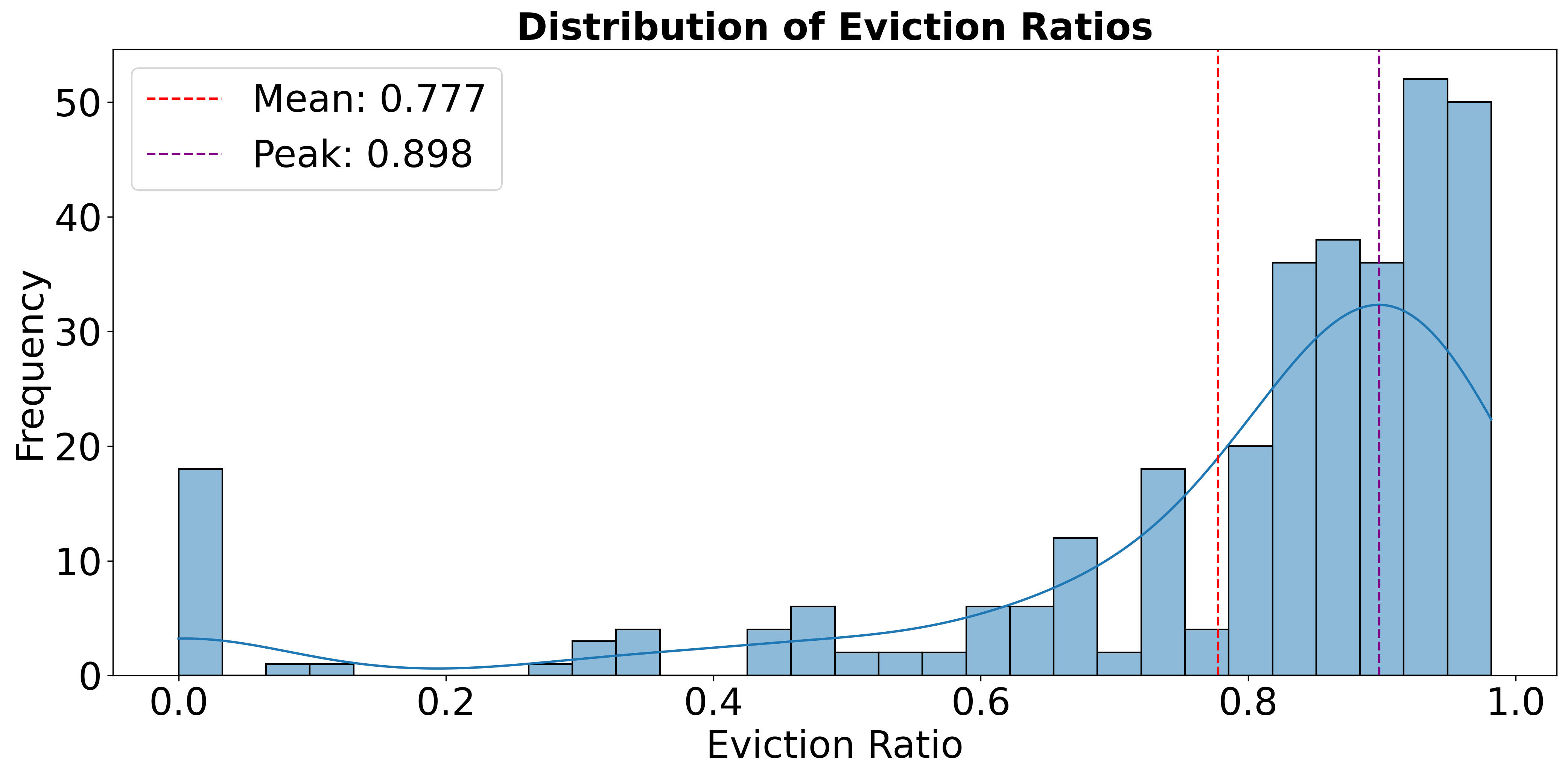}
    \caption{Histograms of FP8 quantization ratios (top) and eviction ratios (bottom) under ARKV. Dashed lines indicate the mean and peak values. Quantization ratios cluster around $\sim$0.14, while eviction ratios vary widely and are strongly influenced by the KV cache budget.}
    \label{fig:ov_mem}
\end{figure}

\textbf{Memory Saving}. Figure~\ref{fig:ov_mem} and Table~\ref{tab:memory_by_cache_size} summarize the memory behavior of ARKV across all models and tasks. The FP8 quantization ratio distribution (top) shows that ARKV consistently applies low-precision storage to only a small portion of tokens, averaging around 14\%. The vertical dashed lines denote the mean and peak of the distribution, and their proximity indicates that quantization usage is both stable and tightly concentrated across runs. This reflects ARKV’s design philosophy: quantization is used conservatively and primarily for medium-importance tokens, rather than as the main driver of memory savings.

In contrast, the eviction ratio distribution (bottom) is broader and strongly influenced by the available KV cache budget. The mean and peak lines appear at higher values, $\sim$0.78 and $\sim$0.90, respectively, showing that the most frequent behavior under constrained budgets is to evict a large fraction of low-impact tokens. As Table VII confirms, eviction ratios decrease as the budget increases, from 0.88 at 512 tokens to 0.66 at 2048 tokens, while the quantization ratio remains nearly constant.

Together, these results indicate that ARKV achieves substantial memory reduction primarily through adaptive eviction, with quantization serving as a stable, budget-independent secondary mechanism. The clear separation between mean and peak in the eviction distribution further highlights the dynamic nature of ARKV’s token selection under different memory constraints.

\begin{table}[htbp]
\centering

\begin{tabular}{lrr}
\toprule
\textbf{Budget} & \textbf{Quant Ratio} & \textbf{Evict Ratio} \\
\hline
512 & 14.39 & 87.80  \\
1024 & 14.41 & 79.47  \\
2048 & 14.47 & 65.97  \\
\bottomrule
\end{tabular}
\caption{Memory Saving by Budget for ARKV (\%)}
\label{tab:memory_by_cache_size}
\end{table}

\section{Conclusion}\label{s:conclusion}

This paper presents ARKV, a tri-state caching framework that dynamically retains, compresses, or evicts key-value cache tokens to reduce memory consumption in LLM inference. Guided by attention-based heavy-hitter scoring and layer-wise $OQ$ scheduling, ARKV efficiently prioritizes both important tokens and sensitive layers, enabling high-quality inference under tight memory budgets. Compared to prior work, 
it offers fine-grained, lightweight, adaptive cache control
by presenting a unified cache management strategy, combining s precision-aware compression and importance-based retention. Experimental results have demonstrated that ARKV achieves up to 4$\times$ memory reduction, $\sim14.4\%$ quantization ratio, with negligible accuracy loss across diverse tasks.

In the future, we plan to extend ARKV with chunked or streaming attention to enable dynamic context extension, where local chunks could utilize compressed caches and outputs can be merged incrementally. We also intend to explore adaptive precision scaling beyond FP8, potentially leveraging hybrid formats or continuous scaling, to achieve tighter control over quantization quality. Finally, generalizing ARKV to Mixture‑of‑Experts (MoE) models and multimodal transformers will broaden its relevance to emerging LLM architectures.

\section*{Acknowledgments}
This work was supported by the GreenDIGIT project (European Union’s Horizon Europe programme, grant No. 101131207).

\bibliographystyle{IEEEtran}
\bibliography{references}

\end{document}